
\magnification \magstep1
\raggedbottom
\openup 4\jot
\voffset6truemm
\headline={\ifnum\pageno=1\hfill\else
\hfill {\it Relativistic gauge conditions in quantum
cosmology} \hfill \fi}
\centerline {\bf RELATIVISTIC GAUGE CONDITIONS}
\centerline {\bf IN QUANTUM COSMOLOGY}
\vskip 0.3cm
\centerline {Giampiero Esposito,$^{1,2}$
Alexander Yu. Kamenshchik,$^{3}$}
\centerline {Igor V. Mishakov,$^{3}$ and
Giuseppe Pollifrone$^{4}$}
\vskip 0.3cm
\centerline {\it ${ }^{1}$Istituto Nazionale di Fisica Nucleare, Sezione di
Napoli,}
\centerline {\it Mostra d'Oltremare Padiglione 20,
80125 Napoli, Italy}
\centerline {\it ${ }^{2}$Dipartimento di Scienze Fisiche,
Mostra d'Oltremare Padiglione 19,
80125 Napoli, Italy}
\centerline {\it ${ }^{3}$Nuclear Safety Institute,
Russian Academy of Sciences,}
\centerline {\it 52 Bolshaya Tulskaya, Moscow 113191, Russia}
\centerline {\it ${ }^{4}$Dipartimento di Fisica, Universit\`a
di Roma ``La Sapienza", and}
\centerline {\it INFN, Sezione di Roma,
Piazzale Aldo Moro 2, 00185 Roma, Italy}
\vskip 0.3cm
\noindent
This paper studies the quantization of the electromagnetic field
on a flat Euclidean background with boundaries. One-loop scaling
factors are evaluated for the one-boundary and two-boundary
backgrounds. The mode-by-mode analysis of Faddeev-Popov quantum
amplitudes is performed by using zeta-function regularization,
and is compared with the space-time covariant evaluation of the
same amplitudes. It is shown that a particular gauge condition
exists for which the corresponding operator matrix acting on
gauge modes is in diagonal form from the beginning. Moreover,
various relativistic gauge conditions are studied
in detail, to investigate the gauge invariance of the perturbative
quantum theory.
\vskip 1cm
\centerline {\bf I. INTRODUCTION}
\vskip 1cm
In a quantum cosmological framework, the quantization of the
electromagnetic field on flat Euclidean backgrounds with
boundaries was first considered in a paper by Louko [1].
The one-loop correction to the Hartle-Hawking wave
function of the universe [2] was studied and the value of $\zeta(0)$
describing the scaling properties of the wave function was calculated
by restricting the path-integral measure to the
physical degrees of freedom (i.e. the transverse part of
the potential). Later, it was found in Ref. [3]
that the value of the scaling factor obtained
by a space-time covariant method on using
the formula for the $A_{2}$ Schwinger-DeWitt coefficient for arbitrary
fields on manifolds with boundaries [4] disagrees with
the result obtained in Ref. [1].
Analogous discrepancies were found for other
higher-spin fields on manifolds with boundaries [5-13], and for gravitons
[14] and photons [15,16] on the Riemannian four-sphere representing
the Wick-rotated version of de Sitter space-time.

Some attempts to understand the reasons of the discrepancies mentioned
above were made in recent years. The first of these ideas suggests
that the reason of discrepancies lies in the
inappropriate implementation of 3+1 split on the manifolds where
this is ill-definite [17].
The second idea is connected with the necessity to
study the contribution of gauge and ghost
modes to the quantum amplitudes [12,18].
The third approach stresses the necessity to pay attention to
a correct definition of the measure in the corresponding path integrals
[15,16,19]. The fourth one consists in the check of the covariant
formulas for the $A_{2}$ Schwinger-DeWitt coefficient for arbitrary
fields on manifolds with boundaries [20].

In our previous paper [21] we have investigated the
correspondence between covariant and non-covariant formalisms for
the Maxwell field on flat Euclidean
four-space with boundaries by applying
the first two approaches mentioned above. We were able to disentangle
the eigenvalue equations for normal and longitudinal components of the
electromagnetic potential $A_{\mu}$ in two relativistic gauges [18,21]:
$$
\Phi_{L} \equiv { }^{(4)}\nabla^{\mu}A_{\mu} \; , \;
\Phi_{E} \equiv { }^{(4)}\nabla^{\mu}A_{\mu}
-A_{0} \; {\rm Tr} \; K \; ,
$$
where $K$ is the extrinsic-curvature tensor of the boundary.
Their contribution to $\zeta(0)$
on the manifold representing the part of flat Euclidean four-space
bounded by two concentric three-spheres was then
evaluated. It was shown that by taking
into account the contributions of non-physical modes and ghosts
(which do not cancel each other in contrast with the usual experience
on the manifolds without curvature or boundaries), one obtains results
for the Faddeev-Popov amplitudes which agree with the space-time
covariant calculation of the same amplitudes.
An analogous result was obtained for
gravitons in the de Donder gauge [22] and for photons in the Coulomb
gauge [23]. Moreover,
it was shown in Ref. [21] that, in the Lorentz gauge,
the value of $\zeta(0)$ on flat Euclidean four-space bounded
by only one three-sphere coincides
with the value of the $A_{2}$ Schwinger-DeWitt
coefficient [24] obtained by using
the corrected formula derived in Ref. [20].

However, relativistic gauges different from the Lorentz gauge
yield a different $\zeta(0)$ value when the boundary
three-geometry consists of only one three-sphere [21].
A possible explanation of these results
can be that the absence of a well-defined 3+1
decomposition of the electromagnetic four-vector potential
on the one-boundary manifold makes the calculations in
terms of physical degrees of freedom and normal and longitudinal
components inconsistent. In the particular case of the Lorentz
gauge, these calculations are still consistent because all the operators
are in relativistically covariant form.

In this paper we continue the analysis of the electromagnetic
field on manifolds with boundaries,
by studying the problem of gauge invariance in Euclidean
Maxwell theory. For this purpose,
we study families of relativistic gauges for
the manifolds with one and two boundaries. In particular,
the gauge is found where the eigenvalue
equations for normal and longitudinal
components are decoupled without having to diagonalize operator
matrices, and the calculations are especially simple.
Since we study a model relevant for the quantization of
closed cosmologies (although in the limiting case of
a flat background [5]),
the normal and tangential components of
the electromagnetic potential are expanded on a family
of three-spheres as [1,12,18,21]
$$
A_{0}(x,\tau)=\sum_{n=1}^{\infty}R_{n}(\tau)Q^{(n)}(x) \; ,
\eqno (1.1)
$$
$$
A_{k}(x,\tau)=\sum_{n=2}^{\infty}\Bigr[f_{n}(\tau)S_{k}^{(n)}(x)
+g_{n}(\tau)P_{k}^{(n)}(x)\Bigr]
\; \; \; \; {\rm for} \; {\rm all} \; k=1,2,3 \; ,
\eqno (1.2)
$$
where $Q^{(n)}(x),S_{k}^{(n)}(x),P_{k}^{(n)}(x)$ are scalar,
transverse and longitudinal vector harmonics on $S^{3}$
respectively [25].

Sec. II shows that a gauge condition exists such that gauge modes
for a spin-1 field can be decoupled without having to use
the diagonalization method described in our previous paper [21].
The resulting $\zeta(0)$ value is obtained.
Sec. III applies the same gauge condition of Sec. II
to flat Euclidean four-space bounded by only one three-sphere.
Sec. IV studies the most general family of relativistic gauge-averaging
functionals depending linearly on gauge modes and their first
derivatives. Results and open problems are presented in Sec. V.
\vskip 1cm
\centerline {\bf II. DECOUPLING OF GAUGE MODES: TWO-BOUNDARY CASE,}
\centerline {\bf  MAGNETIC AND ELECTRIC BOUNDARY CONDITIONS}
\vskip 1cm
Following Refs. [12,18,21], we study quantum amplitudes for Euclidean
Maxwell theory within the framework of Faddeev-Popov formalism.
Thus, the total Euclidean action is given by [12,18]
$$
{\widetilde I}_{E}=I_{gh}+\int_{M}\left[
{1\over 4}F_{\mu \nu}F^{\mu \nu}
+{{\Bigr[\Phi(A)\Bigr]}^{2}\over 2\alpha}\right]
\; \sqrt{{\rm det} \; g} \; d^{4}x  \; ,
\eqno (2.1)
$$
where $A_{\mu}$ is the four-vector potential, $F_{\mu \nu} \equiv
\partial_{\mu}A_{\nu}-\partial_{\nu}A_{\mu}$ denotes the
electromagnetic-field tensor, $g$ is the background four-metric,
$\Phi$ is an arbitrary gauge-averaging functional
defined on a space of connection one-forms, and $\alpha$
is a positive dimensionless parameter. $I_{gh}$ is the corresponding
ghost-field action.

A relevant class of choices for $\Phi(A)$
can be parametrized by a real
number, say $b$, and it can be cast in the form [12,18]
$$
\Phi^{(b)}(A) \; \equiv \; ^{(4)}\nabla^{\mu}\; A_{\mu}-
b \; A_{0} \; {\rm Tr} \; K  \; ,
\eqno (2.2)
$$
where $K$ is the extrinsic-curvature tensor of the boundary.
The two gauges studied in Ref. [21] are a particular case of (2.2),
since $b=0$ leads to the Lorentz gauge,
and $b=1$ yields the Esposito gauge [12,18,21].
If (2.2) is chosen as the gauge-averaging
functional, the part $I_{E}(g,R)$ of the Euclidean action
quadratic in gauge modes is (cf. [12,18])
$$ \eqalignno{
I_{E}(g,R)-{1\over 2\alpha}\int_{\tau_{-}}^{\tau_{+}}\tau^{3}
\Bigr({\dot R}_{1}+{3 \over \tau}(1-b)R_{1}\Bigr)^{2} \; d\tau &=
\sum_{n=2}^{\infty}\int_{\tau_{-}}^{\tau_{+}}
\left[{\tau \over 2(n^{2}-1)}{\Bigr({\dot g}_{n}-(n^{2}-1)R_{n}\Bigr)}^{2}
\right. \cr
& \left.
+{\tau \over 2\alpha}{\Bigr(\tau {\dot R}_{n}+
3(1-b)R_{n}-{g_{n}\over \tau}
\Bigr)}^{2}
\right] d\tau \;  .
&(2.3)\cr}
$$
Of course, we need boundary conditions
on the boundary surfaces.
They can be magnetic, which implies setting to zero on the boundaries
the magnetic field, the gauge-averaging functional and hence the
Faddeev-Popov ghost field. They can also be electric, hence setting
to zero on the boundaries the electric field, and leading to Neumann
conditions on the ghost [12,18,21].
The former imply, in the gauge (2.2), Dirichlet boundary conditions
for $g_{n}$ and ghost modes, and Robin boundary conditions
for $R_{n}$ modes. The latter imply Neumann boundary
conditions for $g_{n}$ and ghost modes, and Dirichlet boundary
conditions for $R_{n}$ modes.

Integrating by parts in (2.3) and using the magnetic or
electric boundary conditions described above
one finds for all $ n \geq 2$
$$ \eqalignno{
I_{E}^{(n)}(g,R) &= {1\over 2}
\int_{\tau_{-}}^{\tau_{+}} {\tau g_{n}\over (n^{2}-1)}
({\widehat A}_{n}g_{n})\; d\tau
+{1\over 2}\int_{\tau_{-}}^{\tau_{+}} \tau^{3}R_{n}
({\widehat B}_{n}R_{n}) \; d\tau \cr
&+\left(1-{1\over \alpha}\right)
\int_{\tau_{-}}^{\tau_{+}}g_{n}{d \over {d\tau}}
(\tau{R}_{n}) \; d\tau
+{3 \over \alpha}
\left(b-{2 \over 3}\right)
\int_{\tau_{-}}^{\tau_{+}}g_{n}R_{n} \; d\tau \; ,
&(2.4)\cr}
$$
where the second-order
elliptic differential operators ${\widehat A}_{n}$
and ${\widehat B}_{n}$ are
$$
{\widehat A}_{n}(\tau) \equiv -{d^{2}\over d\tau^{2}}
-{1\over \tau} {d\over d\tau}
+{(n^{2}-1)\over \alpha \tau^{2}} \; ,
\eqno (2.5)
$$
$$
{\widehat B}_{n}(\tau) \equiv -{1\over \alpha}
\left({d^{2}\over d\tau^{2}}
+{3\over \tau}{d\over d\tau}\right)
+{1 \over \tau^{2}}
\left[n^{2}-1+{3\over \alpha}
\biggr(1+3b\Bigr(b-{4\over 3}\Bigr)\biggr)\right] \; .
\eqno (2.6)
$$
Thus, if we choose the gauge-averaging functional as (cf. (2.2))
$$
\Phi_{P}(A) \equiv { }^{(4)}\nabla^{\mu}A_{\mu}-
{2\over 3} \; A_{0} \; {\rm Tr}
\; K \; ,
\eqno (2.7)
$$
the action quadratic in the gauge
modes becomes for all $n \geq 2$
$$ \eqalignno{
I_{E}^{(n)}(g,R)&={1\over 2}\int_{\tau_{-}}^{\tau_{+}}
-{\tau g_{n}\over (n^{2}-1)}
\biggr[{d^{2}\over d\tau^{2}}+{1\over \tau}{d\over d\tau}
-{(n^{2}-1)\over \alpha \tau^{2}}\biggr]g_{n}\; d\tau \cr
&+{1\over 2}\int_{\tau_{-}}^{\tau_{+}}-\tau^{3}R_{n}
\left[{1\over \alpha}\biggr({d^{2}\over d\tau^{2}}+{3\over \tau}
{d\over d\tau}\biggr)
-\Bigr(n^{2}-1-{1\over \alpha}\Bigr){1\over \tau^{2}}\right]R_{n}
\; d\tau \cr
&+\Bigr(1-{1\over \alpha}\Bigr)
\int_{\tau_{-}}^{\tau_{+}}g_{n}{d \over {d\tau}}
(\tau{R}_{n}) \; d\tau \; .
&(2.8)\cr}
$$
Remarkably, by setting to 1 the parameter $\alpha$
we get the decoupled eigenvalue equations for normal
and longitudinal components of the electromagnetic potential
$$
{{d^{2}g_{n}} \over d\tau^{2}} + {1 \over \tau}\;{{dg_{n}} \over d\tau}
-{(n^{2}-1) \over \tau^{2}}g_{n} + \lambda_{n}g_{n}=0 \; ,
\eqno (2.9)
$$
and
$$
{{d^{2}R_{n}} \over d\tau^{2}} + {3 \over \tau}\;{{dR_{n}} \over d\tau}
-{(n^{2}-2) \over \tau^{2}}R_{n} + \lambda_{n}R_{n}=0 \; .
\eqno (2.10)
$$
The regular solutions of the equations (2.9)-(2.10) are
Bessel functions of non-integer order. However, to use the
complex-contour technique of Refs. [8-11] it is convenient to set
$\lambda_{n}=-M^2$ and then work with the corresponding
modified Bessel functions. After making this change of variable,
and defining $\nu \; \equiv \; \sqrt{n^{2}-1}$, we get
the regular solutions for $g_{n}$ and $R_n$
$$
g_{n}(\tau) = C_{1}\;I_{\nu}({M} \tau)
+ C_{2}\; K_{\nu}(M \tau) \; ,
\eqno (2.11)
$$
$$
R_{n}(\tau)={1 \over \tau}\; \Bigr(C_{3}\;
I_{\nu}(M \tau)
+ C_{4}\;K_{\nu}(M \tau)\Bigr) \; ,
\eqno (2.12)
$$
where $C_{i}$ with $i=1,\cdots,4$ are constants.
As in Ref. [21], both $I$- and $K$-functions contribute to regular
gauge modes, since the singularity at the origin of flat
Euclidean four-space is avoided in our elliptic boundary-value
problem with two three-sphere boundaries.

Now, defining $I_{\nu}^{-} \equiv I_{\nu}(M \tau_{-})$,
$I_{\nu}^{+} \equiv I_{\nu}(M \tau_{+})$,
$K_{\nu}^{-} \equiv K_{\nu}(M \tau_{-})$,
$K_{\nu}^{+} \equiv K_{\nu}(M \tau_{+})$,
and imposing magnetic boundary conditions
described above, one has the following equations:
$$
C_{1}\;I_{\nu}^{-} + C_{2}\;
K_{\nu}^{-}  = 0 \; ,
\eqno (2.13a)
$$
$$
C_{1}\;I_{\nu}^{+} + C_{2}\;
K_{\nu}^{+} = 0 \; ,
\eqno (2.13b)
$$
$$
C_{3}\;
\left(I_{\nu - 1}^{-} + I_{\nu
+ 1}^{-}\right)
- C_{4}\;
\left(K_{\nu - 1}^{-} + K_{\nu
+ 1}^{-}\right)=0 \; ,
\eqno (2.13c)
$$
$$
C_{3}\;
\left(I_{\nu - 1}^{+} + I_{\nu
+ 1}^{+}\right)
- C_{4}\;
\left(K_{\nu - 1}^{+} + K_{\nu
+ 1}^{+}\right)=0 \; .
\eqno (2.13d)
$$
The condition for the existence of non-trivial solutions for the
system (2.13a)-(2.13d) is the vanishing of the determinants
$$
\det\left(\matrix{I_{\nu}^{-}&
K_{\nu}^{-}\cr
I_{\nu}^{+}&
K_{\nu}^{+}\cr}\right)
= 0 \; ,
\eqno (2.14)
$$
$$
\det\left(\matrix{(I_{\nu - 1}^{-} +
I_{\nu +1}^{-})&
-(K_{\nu-1}^{-}+K_{\nu+1}^{-})\cr
(I_{\nu-1}^{+}+I_{\nu+1}^{+})&
-(K_{\nu-1}^{+}+K_{\nu+1}^{+})\cr}\right)
= 0 \; .
\eqno (2.15)
$$
Thus, we have found the condition on eigenvalues for normal and
longitudinal components of the electromagnetic field and can evaluate
their contribution to $\zeta(0)$
by using the algorithm of Refs. [8-11].

For this purpose, let us recall that
$\zeta(0)$ can be expressed as
$$
\zeta(0) = I_{\rm log}+ I_{\rm pole}(\infty)- I_{\rm pole}(0) \; .
\eqno (2.16)
$$
With our notation [8-11], one writes $f_{n}(M^{2})$ for the
function occurring in the equation obeyed by the eigenvalues
by virtue of boundary conditions, and $d(n)$ for the degeneracy
of the eigenvalues. One then defines the function [8-11]
$$
I(M^{2},s) \equiv
\sum_{n=n_{0}}^{\infty} d(n)\;{1 \over n^{2s}}
\; {\rm ln} f_{n}(M^{2}) \; .
\eqno (2.17a)
$$
Such a function has a unique analytic continuation to the
whole complex-$s$ plane as a meromorphic function, i.e.
$$
``I(M^{2},s)"=
{I_{\rm pole}(M^{2})\over s}+I^{R}(M^{2})+{\rm O}(s) \; .
\eqno (2.17b)
$$
Thus, $I_{\log}=I_{\rm log}^{R}$
is the coefficient of ${\rm ln} \; M$ from
$I(M^{2}, s)$ as $M \rightarrow \infty$, and
$I_{\rm pole}(M^{2})$ is the residue at $s=0$.
Remarkably, $I_{\rm log}$ and $I_{\rm pole}(\infty)$
are obtained from uniform asymptotic expansions of
modifies Bessel functions as their order tends to $\infty$
and $M \rightarrow \infty$, whereas $I_{\rm pole}(0)$
is obtained from the limiting behaviour of such Bessel
functions as $M \rightarrow 0$ [8-11].
The condition ${\rm det} \; {\cal I}=0$ (see (2.14)-(2.15))
should be studied after eliminating fake roots
$M = 0$. To obtain that it is enough to divide
${\rm det} \; {\cal I}$ by the minimal
power of M occurring in the determinant. It is easy to
see by using the series expansion for modified Bessel functions
that such a power is $0$ for (2.14) and  $-2$ for (2.15).

We begin with the calculation of $I_{\rm log}$ for $g_n$ and
$R_n$ modes. Using uniform asymptotic expansions of modified
Bessel functions one can see, from (2.14), that the coefficient
of ${\rm ln} \; M$ is $-1$, while (2.15), divided
by $M^{-2}$, gives $+ {\rm ln} \; M$. Hence
$$
I_{\rm log}=I_{{\rm log}_{R_{n}}}
+I_{{\rm log}_{g_{n}}}=0 \; .
\eqno (2.18)
$$
In a similar way one finds that $I_{\rm pole}(0)=0$,
whereas the contributions to
$I_{\rm pole}(\infty)$ from $g_n$ and $R_n$ vanish separately.

The next problem is the calculation of the contribution to $\zeta(0)$
of the $R_1$ mode. In our gauge the eigenvalue equation for it is
$$
{{d^{2}  R_1} \over {d\tau^2}} + {3 \over \tau}{{dR_{1}}\over {d\tau}} +
{{R_1} \over {\tau^2}} - M^{2} R_1 = 0 \; ,
\eqno (2.19)
$$
whose solution is
$$
R_1(\tau)= C_{1}{1 \over \tau}I_{0}(M\tau)
+C_2{1\over \tau}K_{0}(M\tau) \; .
\eqno (2.20)
$$
Imposing Robin (i.e. magnetic) conditions on $R_{1}$
$$
{dR_{1}\over{d\tau}}+ {R_{1}\over \tau}=0 \; ,
\eqno (2.21)
$$
at the three-sphere boundaries, one gets the system of equations
$$
C_{1}I_{1}^{-} - C_{2} K_1^{-} =0 \; ,
\eqno (2.22a)
$$
$$
C_{1}I_{1}^{+} - C_{2} K_1^{+} =0 \; .
\eqno (2.22b)
$$
The determinant of the system (2.22a)-(2.22b) should vanish and this
gives the eigenvalue condition. Such a
determinant has no fake roots. Thus, by using the uniform asymptotic
expansions of Bessel functions one finds that the contribution owed to
$I_{\rm log}$ is $-{1 \over 2}$. As noted in Ref. [21], we have to
add the number $N_{D}=1$ of such decoupled modes to the full
$\zeta(0)$ value. In fact, they are non-trivial since they involve
zero-eigenvalues corresponding to non-vanishing eigenfunctions
[21,26-28]. Hence  one obtains
$$
\zeta_{R_{1}}(0)= I_{\rm log} + N_{D}
= -{1 \over 2} + 1 = {1 \over 2} \; .
\eqno (2.23)
$$

Now we deal with the ghost operator. By studying the gauge
transformation [12,18]
$$
{ }^{\epsilon}A_{\mu}
\equiv A_{\mu}+{ }^{(4)}\nabla_{\mu}\epsilon \; ,
\eqno (2.24)
$$
one gets, by virtue of (2.2),
$$
\Phi^{(b)}(A)-\Phi^{(b)}({ }^{\epsilon}A)
=\sum_{n=1}^{\infty} \left[-{{d^{2}} \over {d\tau^{2}}}
-{3 \over \tau}(1-b){d \over {d\tau}}
+{{(n^{2}-1)}\over {\tau^{2}}}\right]\epsilon_{n}{(\tau)}
Q^{(n)}(x) \; .
\eqno (2.25)
$$
Hence the eigenfunctions of the ghost operator are
related to [12,18]
$$
\epsilon_{n}({\tau}) = \tau^{({{3b} \over {2}} -1)}
\Bigr(B_{1}I_{\nu}(M{\tau})+B_{2}K_{\nu}(M{\tau})\Bigr) \; ,
\eqno(2.26)
$$
where
$$
\nu \equiv +\sqrt{n^{2}-{3b \over 4}(4-3b)} \; .
\eqno (2.27)
$$
In the case  $b={2 \over 3}$, the order of the modified Bessel
functions in (2.26) is $+\sqrt{n^{2}-1}$ as in (2.11)-(2.12).
The contribution to $\zeta(0)$ of the ghost, in both cases (magnetic
and electric) is zero. Bearing in mind
that, from Ref. [21], the contribution to $\zeta(0)$
of transverse modes is $-{1 \over 2}$ with magnetic
boundary conditions, and $1 \over 2$ when the boundary
conditions are electric, one gets
$$
\zeta(0)=\zeta_{ \rm transversal \; photons}(0)
+ \zeta_{R_{1}}(0)={-{1 \over 2}} +
{1 \over 2} = 0 \; .
\eqno (2.28)
$$

The calculation of $\zeta(0)$ in the electric case is immediate.
In this case ${\dot g}_{n}=0$ and $R_{n}=0$ at the three-sphere boundaries,
and only the decoupled mode contributes to the $\zeta(0)$ value
and it yields $\zeta_{R_{1}}(0)=-{1 \over 2}$.
Thus, also in this case, one obtains $\zeta(0)=0$.

Our results coincide with those obtained by a space-time covariant
Schwinger-DeWitt method, where the vanishing of the $A_{2}$
coefficient results from the mutual cancellation of the contributions
from the two boundaries, in the case of flat Euclidean
four-space [24].
\vskip 1cm
\centerline {\bf III. DECOUPLING OF GAUGE MODES}
\centerline {\bf IN THE ONE-BOUNDARY PROBLEM}
\vskip 1cm
Since the gauge condition studied in the previous section leads
more easily to the decoupling of $g_n$ and $R_n$ modes, and
it agrees with the results found in Ref. [21] in the two-boundary case,
it appears necessary to study its properties in the one-boundary
problem as well. Moreover, this analysis enables one to further check
the gauge dependence of the one-loop quantum amplitudes [21].
Following the results of Sec. II, we can write the regular
solution for $g_n$, $R_n$ and $\epsilon_n$ as
$$
g_{n}(\tau)=A\; I_{\nu}(M\tau) \; ,
\eqno(3.1)
$$
$$
R_{n}(\tau)=B\;{ 1 \over \tau}I_{\nu}(M\tau) \; ,
\eqno(3.2)
$$
$$
{\epsilon_{n}({\tau})}=C \; I_{\nu}(M\tau) \; ,
\eqno(3.3)
$$
where $A, B, C$ are constants.
Imposing magnetic boundary conditions at the
three-sphere boundary of radius $a$,
we get
$$
{I_{\nu}{(Ma)}}=0
\eqno(3.4)
$$
for $g_n$ and $\epsilon_n$, and
$$
{I'_{\nu}{(Ma)}}=0
\eqno(3.5)
$$
for $R_n$. Remarkably, the only possible form of the
decoupled mode for normal photons is $R_{1} \equiv 0$,
since $R_1$ would be proportional to $I_{0}(M{\tau})/\tau$
in our gauge, and hence cannot be regular at the origin
(see also the end of Sec. IV).

First, we evaluate $I_{ \rm log}$ for $g_n$ and $\epsilon_n$.
Using, as usual, the uniform asymptotic
expansion of modified Bessel functions,
eliminating fake roots $M=0$ and
taking into account that the degeneracy of ghost modes is $-2$
times the degeneracy of $g_n$, we see that the coefficient of
${\rm ln} M$ is $\nu + {1 \over 2}$, where
$M^{\nu}$ is the power of fake roots. For $R_n$, after dividing
by $M^{\nu-1}$, we find that
the coefficient of ${\rm ln} M$ is $-(\nu - {1 \over 2})$.
Hence we obtain
$$
I_{\rm log}={\sum_{n=2}^{\infty}}{{n^2} \over 2}= -{1 \over 2} \; .
\eqno(3.6)
$$
For $\epsilon_1$, which is proportional to $I_{0}(M\tau)$,
we get by a simple calculation
$$
{{I_{\rm log}}_{\epsilon_1}}
={1\over 2}  \; .
\eqno(3.7)
$$

It is easy to see that the contribution
to $I_{\rm pole}(\infty)$ is equal to zero
for $g_{n}$, $R_{n}$ and $\epsilon_{n}$ separately.
Last, we have to evaluate
$I_{\rm pole}(0)$. The contribution of
$g_{n}$ to $I_{\rm pole}(0)$ is obtained by taking
the coefficient of ${1\over n}$
in the asymptotic expansion as $n \rightarrow \infty$ of
$$
{1\over 2}n^{2} {\rm ln}  {1\over {\Gamma(\nu+1)}} \; ,
$$
while the structure of the term deriving from $R_{n}$ which contributes to
$I_{\rm pole}(0)$ is
$$
{1 \over 2}{n^2}
{\rm ln} {1\over \Gamma(\nu)} \; .
$$
Both terms contribute $-{59\over 720}$ to $I_{\rm  pole}(0)$, and
bearing in mind the different degeneracy between
ghost and gauge modes one finds
$$
I_{\rm pole}(0)
=  0 \;.
\eqno(3.8)
$$
Finally, taking into account the contribution to
$\zeta(0)$ of the transverse part of the potential [1]
we get the full $\zeta(0)$ as
$$
\zeta(0)=-{77 \over 180} \; .
\eqno(3.9)
$$
Remarkably, this $\zeta(0)$ value agrees with the one obtained
in Ref. [1], where ghost and gauge modes were not taken
into account. The striking cancellation of $\zeta(0)$
contributions from ghost and gauge modes {\it in the particular
gauge} (2.7) deserves further thinking.

If we choose electric boundary conditions at the three-sphere
boundary, the roles of $g_n$
and $R_n$ are interchanged and ghost
modes obey Neumann boundary conditions. Hence, a similar
analisys leads to the full $\zeta(0)$ value [12,18]
$$
\zeta(0)={13 \over 180}+ {1\over 2}+{3\over 2}+1
= {553 \over 180} \; .
\eqno (3.10)
$$

The results (3.9)-(3.10) show that,
on choosing the gauge condition (2.7) in the one-boundary
problem, the full $\zeta(0)$ value
is different on imposing magnetic or
electric boundary conditions. However, an analysis along
the lines of Ref. [21] and of this section shows that, on
imposing electric boundary conditions in the Lorentz gauge,
one finds again $\zeta(0)=-{31\over 90}$ as in Ref. [21],
where magnetic boundary conditions were studied in the
one-boundary problem. The dependence of the $\zeta(0)$ value
on the boundary conditions and on the gauge conditions seems
to result from the ill-definite nature of the 3+1 split of
our background with only one boundary (see Sec. V).
\vskip 1cm
\centerline {\bf IV. THE MOST GENERAL GAUGE-AVERAGING FUNCTIONAL}
\vskip 1cm
In Refs. [12,18,21] and in Sec. II of this
paper, gauge invariance of the
Faddeev-Popov formalism in the presence of boundaries has been
{\it assumed} to obtain a convenient set of eigenvalue equations
leading to the full $\zeta(0)$ value for one-loop quantum
amplitudes. To complete our analysis it is therefore necessary
to study the most general gauge-averaging functional $\Phi(A)$.
Of course, the family (2.2) of gauge functionals is only a
{\it particular} case. Our $\Phi(A)$ should obey the following
conditions:

(i) $\Phi(A)$ is linear in the gauge modes and their first
derivatives, to ensure that the total Euclidean action is
quadratic in the gauge modes and only involves second-order
elliptic operators.

(ii) $\Phi(A)$ does not contain first derivatives of $g_{n}$ modes.
In fact, such derivatives only occur in the components of the
electric field, but not in the Lorentz functional, or in the
Coulomb functional, or in the $A_{0} \; {\rm Tr} \; K$ term.
Moreover, the variation of the total Euclidean action does not
vanish if the contribution of ${\dot g}_{n}$ is added to
$\Phi(A)$.
\vskip 0.3cm
\noindent
One is thus led to write $\Phi(A)$ in the form
$$ \eqalignno{
\Phi(A) & \equiv \biggr(\gamma_{1}{\dot R}_{1}+\gamma_{2}
{R_{1}\over \tau} \biggr)Q^{(1)}(x)
+\sum_{n=2}^{\infty}\biggr(\gamma_{1}{\dot R}_{n}
+\gamma_{2}{R_{n}\over \tau}+\gamma_{3}{g_{n}\over \tau^{2}}
\biggr) Q^{(n)}(x) \cr
&= \gamma_{1} { }^{(4)}\nabla^{0}A_{0}
+{1\over 3}\gamma_{2} \; A_{0} \; {\rm Tr} \; K
-\gamma_{3} \; { }^{(3)}\nabla^{i}A_{i} \; ,
&(4.1)\cr}
$$
where $\gamma_{1},\gamma_{2},\gamma_{3}$ are arbitrary
dimensionless parameters independent of $\tau$.
Note that, if $\gamma_{1}$ does not vanish, it can be absorbed
into the definition of $\alpha$ by setting
${\gamma_{1}^{2}\over \alpha} \equiv {1\over {\widetilde \alpha}}$,
whereas ${\gamma_{2}\over \gamma_{1}} \equiv {\widetilde \gamma}_{2}$,
${\gamma_{3}\over \gamma_{1}} \equiv {\widetilde \gamma}_{3}$. This
imples that, if $\gamma_{1} \not = 0$, one can always consider an
equivalent quantum theory where $\gamma_{1}=1$, while $\gamma_{2}$
and $\gamma_{3}$ remain arbitrary. An equivalent classification is
obtained by focusing on $\gamma_{2}$ or $\gamma_{3}$. With this
understanding, the following (sub)families of non-vanishing
gauge functionals may occur:
(1) $\gamma_{1}=1$, $\gamma_{2} \not = 0$, $\gamma_{3}
\not =0$;
(2) $\gamma_{1}=1$, $\gamma_{2}=0$, $\gamma_{3} \not =0$;
(3) $\gamma_{1}=1$, $\gamma_{2} \not =0$, $\gamma_{3}=0$;
(4) $\gamma_{1}=1$, $\gamma_{2}=\gamma_{3}=0$;
(5) $\gamma_{1}=0$, $\gamma_{2} \not =0$, $\gamma_{3}
\not =0$;
(6) $\gamma_{1}=\gamma_{2}=0$, $\gamma_{3} \not =0$;
(7) $\gamma_{1}=0$, $\gamma_{2} \not =0$, $\gamma_{3}=0$.

The cases (5)-(7) correspond to {\it degenerate} gauge
functionals, in that they do not lead to second-order elliptic
operators on $R_{n}$ modes. They are not studied
in this paper (cf. [23]). Hence we here
focus on the cases (1)-(4), i.e. whenever $\gamma_{1}$ does not
vanish (see above). The first problem we face is the attempt to
decouple $g_{n}$ and $R_{n}$ modes by means of the operator
matrix first applied in Ref. [21]. In our case, by virtue of (2.1)
and (4.1), the coupled eigenvalue equations take the form
(cf. [21])
$$
{\widehat A}_{n}g_{n}(\tau)+{\widehat B}_{n}R_{n}(\tau)=0 \; ,
\eqno (4.2)
$$
$$
{\widehat C}_{n}g_{n}(\tau)+{\widehat D}_{n}R_{n}(\tau)=0 \; ,
\eqno (4.3)
$$
where, on defining
$$
\rho \equiv 1+{\gamma_{3}\over \alpha} \; ,
\eqno (4.4)
$$
$$
\mu \equiv 1+ {\gamma_{2}\gamma_{3}\over \alpha} \; ,
\eqno (4.5)
$$
one has
$$
{\widehat A}_{n} \equiv {d^{2}\over d\tau^{2}}
+{1\over \tau}{d\over d\tau}-{\gamma_{3}^{2}\over \alpha}
{(n^{2}-1)\over \tau^{2}}+\lambda_{n} \; ,
\eqno (4.6)
$$
$$
{\widehat B}_{n} \equiv -\rho (n^{2}-1){d\over d\tau}
-{\mu (n^{2}-1)\over \tau} \; ,
\eqno (4.7)
$$
$$
{\widehat C}_{n} \equiv {\rho \over \tau^{2}}{d\over d\tau}
+{\gamma_{3}\over \alpha}(1-\gamma_{2}){1\over \tau^{3}} \; ,
\eqno (4.8)
$$
$$
{\widehat D}_{n} \equiv {1\over \alpha}{d^{2}\over d\tau^{2}}
+{3\over \alpha}{1\over \tau}{d\over d\tau}
+\biggr[{\gamma_{2}\over \alpha}(2-\gamma_{2})-(n^{2}-1)\biggr]
{1\over \tau^{2}}+\lambda_{n} \; .
\eqno (4.9)
$$
As we did in Ref. [21], we now look for a diagonalized matrix in
the form
$$
O_{ij}^{(n)}  \equiv
\pmatrix {1& V_{n}(\tau)\cr W_{n}(\tau)&1 \cr}
\pmatrix {{\widehat A}_{n}& {\widehat B}_{n}\cr
{\widehat C}_{n}& {\widehat D}_{n}\cr}
\pmatrix {1& \alpha_{n}(\tau)\cr \beta_{n}(\tau)&1 \cr} \; .
\eqno (4.10)
$$
Thus, on using the operator identities [21]
$$
\biggr[{d\over d\tau},\alpha_{n}\biggr]
={d\alpha_{n}\over d\tau} \; ,
\eqno (4.11)
$$
$$
\biggr[{d^{2}\over d\tau^{2}},\alpha_{n}\biggr]
={d^{2}\alpha_{n}\over d\tau^{2}}+2{d\alpha_{n}\over d\tau}
{d\over d\tau} \; ,
\eqno (4.12)
$$
and setting to zero the off-diagonal matrix element
$$
O_{12}^{(n)}=
{\widehat A}_{n}\alpha_{n}+{\widehat B}_{n}
+V_{n}{\widehat C}_{n}\alpha_{n}
+V_{n}{\widehat D}_{n} \; ,
$$
one finds the system of equations (cf. [21])
$$
V_{n}+\alpha_{n}=0 \; ,
\eqno (4.13)
$$
$$
2{d\alpha_{n}\over d\tau}+2\Bigr(1-{1\over \alpha}\Bigr)
{dV_{n}\over d\tau}
+{\Bigr(\alpha_{n}+3{V_{n}\over \alpha}\Bigr)\over \tau}
-\rho (n^{2}-1) =0 \; ,
\eqno (4.14)
$$
$$ \eqalignno{
\; & {d^{2}\alpha_{n}\over d\tau^{2}}
+\biggr({\rho V_{n}\over \tau^{2}}+{1\over \tau}\biggr)
{d\alpha_{n}\over d\tau}-{\gamma_{3}^{2}\over \alpha}
{(n^{2}-1)\over \tau^{2}}\alpha_{n}-(n^{2}-1){\mu \over \tau} \cr
&+{\gamma_{3}\over \alpha}\Bigr(1-\gamma_{2}\Bigr)
V_{n}\alpha_{n}{1\over \tau^{3}}
+\biggr[{\gamma_{2}\over \alpha}(2-\gamma_{2})-(n^{2}-1)\biggr]
{V_{n}\over \tau^{2}}=0 \; .
&(4.15)\cr}
$$
Equations (4.13)-(4.14) are solved by $V_{n}=-\alpha_{n}$,
and
$$
\alpha_{n}(\tau)={\alpha \over (\alpha-1)}\rho (n^{2}-1)\tau
+\alpha_{0,n}\tau^{(3-\alpha)\over 2} \; ,
\eqno (4.16)
$$
where $\alpha_{0,n}$ is a constant. Since the insertion of
(4.16) into (4.15) leads to an involved condition unless
$\alpha=1$, it is very interesting to study first the limiting
case $\alpha \rightarrow \infty$. This does not affect the
arbitrariness in the choice of the parameters
$\gamma_{1},\gamma_{2},\gamma_{3}$ appearing in (4.1). One
then finds the condition
$$ \eqalignno{
\; & \biggr(1+{\gamma_{3}\over \alpha}(2-\gamma_{2})\biggr)
{\alpha^{2}\over (\alpha-1)^{2}}(n^{2}-1)^{2}
\Bigr(1+{\gamma_{3}\over \alpha}\Bigr)^{2} \cr
&+\biggr[{\gamma_{3}^{2}\over \alpha}(n^{2}-1)
+{\gamma_{2}\over \alpha}(2-\gamma_{2})-n^{2}\biggr]
{\alpha \over (\alpha-1)}(n^{2}-1)
\Bigr(1+{\gamma_{3}\over \alpha}\Bigr) \cr
&=-(n^{2}-1)\Bigr(1+{\gamma_{2}\gamma_{3}\over \alpha}\Bigr) \; ,
&(4.17)\cr}
$$
which is identically satisfied for all $n \geq 2$, as
$\alpha \rightarrow \infty$. This shows that the limiting
form of $\alpha_{n}(\tau)$ as $\alpha \rightarrow \infty$,
i.e.
$$
\alpha_{n}(\tau) \sim (n^{2}-1)\tau \; ,
\eqno (4.18)
$$
is indeed also a solution of equation (4.15).

One now has to set to zero the off-diagonal matrix element
$$
O_{21}^{(n)}=W_{n}{\widehat A}_{n}+W_{n}{\widehat B}_{n}\beta_{n}
+{\widehat C}_{n}+{\widehat D}_{n}\beta_{n} \; ,
$$
in (4.10). By virtue of (4.6)-(4.9),
and (4.11)-(4.12) applied to $\beta_{n}(\tau)$, one thus
finds the system of equations
$$
W_{n}+\beta_{n}=0 \; ,
\eqno (4.19)
$$
$$
2{d\beta_{n}\over d\tau}+{\Bigr(W_{n}+{3\over \alpha}\beta_{n}\Bigr)
\over \tau}+{\rho \over \tau^{2}}=0 \; ,
\eqno (4.20)
$$
$$ \eqalignno{
\; & {1\over \alpha}{d^{2}\beta_{n}\over d\tau^{2}}
+\biggr({3\over \alpha}{1\over \tau}-\rho(n^{2}-1)W_{n}
\biggr){d\beta_{n}\over d\tau}
-\mu(n^{2}-1)W_{n} \; \beta_{n} \; {1\over \tau} \cr
&+\left[\biggr({\gamma_{2}\over \alpha}(2-\gamma_{2})
-(n^{2}-1)\biggr)\beta_{n}-{\gamma_{3}^{2}\over \alpha}
(n^{2}-1)W_{n}\right]{1\over \tau^{2}}
+{\gamma_{3}\over \alpha}(1-\gamma_{2}){1\over \tau^{3}}=0 \; .
&(4.21)\cr}
$$
Equation (4.19) implies $W_{n}=-\beta_{n}$. Hence (4.20) is
solved by
$$
\beta_{n}(\tau)
={\alpha \over (\alpha-1)}{\rho \over 3\tau}
+\beta_{0,n} \; \tau^{{1\over 2}(1-{3\over \alpha})} \; ,
\eqno (4.22)
$$
where $\beta_{0,n}$ is a constant. However, a direct calculation
shows that the limiting form of $\beta_{n}(\tau)$ as
$\alpha \rightarrow \infty$, i.e.
$$
\beta_{n}(\tau) \sim {1\over 3\tau}
+ \beta_{0,n}\sqrt{\tau} \; ,
\eqno (4.23)
$$
is not a solution of (4.21) as $\alpha \rightarrow \infty$.
Moreover, if one studies finite values of $\alpha$, the
exact formulae (4.16) and (4.22), on insertion into (4.15)
and (4.21) respectively, lead to equations which are not
satisfied unless the parameters $\gamma_{2},\gamma_{3}$ and
$\alpha$ take very special values. For example, if
$\gamma_{2}=1,\gamma_{3}=-\alpha=-1$, $\alpha_{n}=\beta_{n}
=V_{n}=W_{n}=0$, the decoupling functional (2.7) is recovered.

Thus, our analysis shows that gauge modes cannot be decoupled
for arbitrary gauge-averaging functionals, and one now faces the
problem of evaluating their contribution to the full $\zeta(0)$
even though $g_{n}$ and $R_{n}$ are not expressed in terms of
Bessel functions [12,18,21,29]. However, for the class of gauge
conditions (2.2) involving the arbitrary dimensionless parameter $b$,
the basis functions can be found by using the technique
described in Ref. [21].
The resulting $\zeta(0)$ value in the two-boundary
problem is again equal to zero for magnetic or electric
boundary conditions, while in the one-boundary problem
the $\zeta(0)$ value depends on $b$. In the case of
magnetic boundary conditions one finds
$$ \eqalignno{
\zeta_{b}(0)&=-{8\over 45}-{1\over 96}\Bigr(3b-2\Bigr)
\Bigr(27 b^{3}-36 b^{2} -12 b -8\Bigr) \cr
&+\Bigr({{\mid 3b -2 \mid}\over 2} -{1\over 4} \Bigr)
\Bigr(1-\theta(b-1)\Bigr)\biggr(1-\theta \Bigr({1\over 3}-b\Bigr)
\biggr) \; .
&(4.24)\cr}
$$
Note that the last term in Eq. (4.24)
reflects the absence of a regular decoupled mode $R_{1}$
for $b \in ]{1\over 3}, 1[$, in agreement with what
we found in the particular case of Sec. III. One can
easily check that Eq. (4.24) agrees with the $\zeta(0)$
values obtained in Ref. [21] and in our Sec. III.
\vskip 10cm
\centerline {\bf V. RESULTS AND OPEN PROBLEMS}
\vskip 1cm
In this paper we have obtained the following results.
First, we have studied the class of gauge functionals
for which the disentanglement of the eigenvalue equations
for normal and longitudinal modes can be achieved, and
we have pointed out one particular choice when such
equations are decoupled from the beginning.
Second, on using this particular gauge functional, the
calculation of the full $\zeta(0)$ value,
in the two-boundary problem,
agrees with the evaluation
performed in Ref. [21],
where we have imposed other gauge conditions.
Third, in the one-boundary problem, we have found that
the one-loop quantum amplitudes are gauge-dependent and
the computation of the full $\zeta(0)$ value is different on
imposing magnetic or electric boundary conditions.
These undesirable properties,
as already noted in Refs. [17,21], seem to add
evidence in favour of the 3+1 decomposition of the
four-vector potential being ill-definite on the manifolds
bounded by only one three-surface. Fourth, we have studied
the most general class of relativistic gauges and the
corresponding eigenvalue equations have been obtained for
the first time.

Interestingly, the recent work in the literature shows that the
semiclassical amplitudes respect the properties of the underlying
classical theory. For example, for a massless spin-${1\over 2}$
field obeying the Weyl equation and subject to spectral or locally
supersymmetric boundary conditions on a three-sphere, the regular
modes turn out to obey the same boundary conditions [12,30].
In the one-loop quantum theory, the eigenvalue conditions are
different, but the $\zeta(0)$ values turn out to coincide
[6,7,9-12]. Moreover, Euclidean Maxwell theory in vacuum is invariant
under duality transformations. Correspondingly, we have found that the
one-loop amplitudes are independent of the choice of electric or
magnetic boundary conditions, providing the Lorentz gauge is chosen
in the one-boundary problem.

The main open problem in Euclidean Maxwell theory in the
presence of boundaries seems to be the {\it explicit}
proof of gauge invariance of one-loop amplitudes for
relativistic gauges, in the case of flat Euclidean space
bounded by two concentric three-spheres. For this purpose,
one may have to show that, for coupled gauge modes,
$I_{\rm log}$ and the difference $I_{\rm pole}(\infty)
-I_{\rm pole}(0)$ are not affected by a change in the gauge
parameters $\gamma_{1},\gamma_{2},\gamma_{3},\alpha$ of
Sec. IV. Although this is what happens in the particular
cases studied so far, at least three technical achievements are
necessary to obtain a rigorous proof, i.e.
\vskip 0.3cm
\noindent
(i) To relate the regularization at large $x$ used in Refs.
[12,18] to the regularization based on the BKKM function
defined in (2.17a).
\vskip 0.3cm
\noindent
(ii) To evaluate $I_{\rm log}$ from an asymptotic analysis
of coupled eigenvalue equations.
\vskip 0.3cm
\noindent
(iii) To evaluate $I_{\rm pole}(\infty)-I_{\rm pole}(0)$ by
relating the analytic continuation to the whole complex-$s$
plane of the difference $I(\infty,s)-I(0,s)$ (see (2.17a)) to
the analytic continuation of the zeta-function.

If this last step can be performed, it may involve an
integral transform relating the BKKM function (2.17a) to the
zeta-function, and a non-trivial application of the
Atiyah-Patodi-Singer theory of Riemannian four-manifolds with
boundary [26,31]. In other words, one might have to prove
that, {\it in the two-boundary problem only},
$I_{\rm pole}(\infty)-I_{\rm pole}(0)$ resulting from coupled
gauge modes is the residue of a meromorphic function, invariant
under a smooth variation in $\gamma_{1},\gamma_{2},\gamma_{3},
\alpha$ of the matrix of elliptic self-adjoint operators appearing
in (4.6)-(4.9). Work is now in progress on this problem, and
we hope to be able to solve it in a future publication.

There is also the problem of physical interpretation of
the results obtained so far [18,21]. In the two-boundary case,
where one has a well-defined 3+1 split of the electromagnetic
potential, the contributions to $\zeta(0)$ which, jointly
with transverse modes, enable one to obtain agreement
with the space-time covariant calculation, result only from
the decoupled gauge modes. Note that such decoupled modes
should be treated separately, since they do not correspond
to any Dirac constraint of the theory [19]. However, in
the case of flat Euclidean space bounded by only one three-sphere,
even on studying the Lorentz gauge which leads to agreement
between mode-by-mode and space-time covariant calculations
of Faddeev-Popov amplitudes, the non-vanishing contributions
to $\zeta(0)$ are not due just to transverse modes and
decoupled modes. By contrast, longitudinal, normal
and ghost modes play a role as well in obtaining the
full $\zeta(0)$ value. Perhaps, the re-definition of the
very notion of physical degrees of freedom is necessary in
this case, and the problem deserves further consideration.
\vskip 1cm
\centerline {\bf ACKNOWLEDGMENTS}
\vskip 1cm
Our research was partially supported by the European Union
under the Human Capital and Mobility Program.
Moreover, the research described in this publication was made
possible in part by Grant No MAE000 from the International
Science Foundation. The work of A. Yu. Kamenshchik was partially
supported by the Russian Foundation for Fundamental Researches
through grant No 94-02-03850-a, and by the Russian
Research Project ``Cosmomicrophysics". The first two
authors enjoyed many conversations with leading experts
during the Heat-Kernel Conference at Winnipeg in August 1994.
\vskip 1cm
\hrule
\vskip 1cm
\item {[1]}
J. Louko, Phys. Rev. D {\bf 38}, 478 (1988).
\item {[2]}
J.B. Hartle and S.W. Hawking, Phys. Rev. D {\bf 28}, 2960 (1983);
S.W. Hawking, Nucl. Phys. {\bf B239}, 257 (1984).
\item {[3]}
I.G. Moss and S. Poletti, Nucl. Phys. {\bf B341}, 155 (1990).
\item {[4]}
T.P. Branson and P.B. Gilkey, Commun. Part. Diff.
Eq. {\bf 15}, 245 (1990).
\item {[5]}
K. Schleich, Phys. Rev. D {\bf 32}, 1889 (1985).
\item {[6]}
P.D. D'Eath and G. Esposito,
Phys. Rev. D {\bf 43}, 3234 (1991).
\item {[7]}
P.D. D'Eath and G. Esposito,
Phys. Rev. D {\bf 44}, 1713 (1991).
\item {[8]}
A.O. Barvinsky, A.Yu. Kamenshchik, and I.P.
Karmazin, Ann. Phys. (N. Y.) {\bf 219}, 201 (1992).
\item {[9]}
A.O. Barvinsky, A.Yu. Kamenshchik, I.P. Karmazin,
and I.V. Mishakov, Class. Quantum Grav. {\bf 9}, L27 (1992).
\item {[10]}
A.Yu. Kamenshchik and I.V. Mishakov,
Int. J. Mod. Phys. A {\bf 7}, 3713 (1992).
\item {[11]}
A.Yu. Kamenshchik and I.V. Mishakov,
Phys. Rev. D {\bf 47}, 1380 (1993).
\item {[12]}
G. Esposito, {\it Quantum Gravity, Quantum
Cosmology and Lorentzian Geometries},
Lecture Notes in Physics: Monographs, Vol. m 12
(Springer-Verlag, Berlin, 1994).
\item {[13]}
G. Esposito, Int. J. Mod. Phys. D {\bf 3}, 593 (1994).
\item {[14]}
P.A. Griffin and D.A. Kosower, Phys. Lett.
B {\bf 233}, 295 (1989).
\item {[15]}
D.V. Vassilevich, Nuovo Cimento A {\bf 104}, 743 (1991).
\item {[16]}
D.V. Vassilevich, Nuovo Cimento A {\bf 105}, 649 (1992).
\item {[17]}
A.Yu. Kamenshchik and I.V. Mishakov, Phys. Rev. D {\bf 49}, 816 (1994).
\item {[18]}
G. Esposito, Class. Quantum Grav. {\bf 11}, 905 (1994).
\item {[19]}
D.V. Vassilevich, {\it QED on Curved Background and on Manifolds with
Boundaries: Unitarity versus Covariance}, ICTP preprint IC/94/359,
Trieste, 1994 (unpublished).
\item {[20]}
D.V. Vassilevich, {\it Vector Fields on a Disk with Mixed Boundary
Conditions}, St. Petersburg preprint SPbU-IP-94-6,
to appear in J. Math. Phys (unpublished).
\item {[21]}
G. Esposito, A.Yu. Kamenshchik, I.V. Mishakov, and G. Pollifrone,
Class. Quantum Grav. {\bf 11}, 2939 (1994).
\item {[22]}
G. Esposito, A.Yu. Kamenshchik, I.V. Mishakov, and G. Pollifrone,
Phys. Rev. D {\bf 50}, 6329 (1994).
\item {[23]}
G. Esposito and  A.Yu. Kamenshchik, Phys. Lett. B {\bf 336}, 324
(1994).
\item {[24]}
I.G. Moss and S. Poletti, Phys. Lett. B {\bf 333}, 326 (1994).
\item {[25]}
E.M. Lifshitz and I.M. Khalatnikov, Adv. Phys. {\bf12}, 185 (1963).
\item {[26]}
M.F. Atiyah, V.K. Patodi, and I.M. Singer, Math. Proc. Camb.
Philos. Soc. {\bf 79}, 71 (1976).
\item {[27]}
S.M. Christensen and M.J. Duff, Nucl. Phys. {\bf B170},
480 (1980).
\item {[28]}
E.S. Fradkin and A.A. Tseytlin, Nucl. Phys. {\bf B234},
472 (1984).
\item {[29]}
G. Esposito, Nuovo Cimento B {\bf 109}, 203 (1994).
\item {[30]}
G. Esposito, H.A. Morales-T\'ecotl, and G. Pollifrone,
Found. Phys. Lett. {\bf 7}, 303 (1994).
\item {[31]}
G. Esposito, in {\it Proceedings of the Conference on
Heat-Kernel Techniques and Quantum Gravity, Winnipeg},
edited by S. Fulling (Winnipeg, Canada, 1994).

\bye